# Effects of Capping on the Ga$_{1-x}$Mn$_x$As Magnetic Depth Profile

B. J. Kirby
Department of Physics and Astronomy, University of Missouri,
Columbia, Missouri  65211, USA

J. A. Borchers
NIST Center for Neutron Research, National Institute of Standards and Technology,
Gaithersburg, Maryland  20899, USA

J. J. Rhyne
Manjuel Lujan Jr. Neutron Scattering Center, Los Alamos National Laboratory,
Los Alamos, New Mexico  87545, USA and
Department of Physics and Astronomy, University of Missouri,
Columbia, Missouri  65211, USA

K. V. O'Donovan
NIST Center for Neutron Research, National Institute of Standards and Technology,
Gaithersburg, Maryland 20899, USA and
Department of Physiology and Biophysics, University of California, Irvine, California
92697

T. Wojtowicz
Department of Physics, University of Notre Dame, Notre Dame, Indiana  46556, USA
and
Institute of Physics of the Polish Academy of Sciences, 02-688 Warsaw, Poland

X. Liu, Z. Ge, S. Shen, and J. K. Furdyna
Department of Physics, University of Notre Dame, Notre Dame, Indiana  46556, USA

Annealing can increase the Curie temperature and net magnetization in uncapped $Ga_{1-x}Mn_xAs$ films, effects that are suppressed when the films are capped with GaAs. Previous polarized neutron reflectometry (PNR) studies of uncapped $Ga_{1-x}Mn_xAs$ revealed a pronounced magnetization gradient that was reduced after annealing. We have extended this study to $Ga_{1-x}Mn_xAs$ capped with GaAs. We observe no increase in Curie temperature or net magnetization upon annealing. Furthermore, PNR measurements indicate that annealing produces minimal differences in the depth-dependent magnetization, as both as-grown and annealed films feature a significant magnetization gradient. These results suggest that the GaAs cap inhibits redistribution of interstitial Mn impurities during annealing.

The emerging field of "spintronics" has motivated recent interest in developing high Curie temperature ($T_C$) ferromagnetic semiconductors. $Ga_{1-x}Mn_xAs$ is a possible candidate for spintronic applications, with a maximum achieved $T_C \approx$ 150 K.[1,2] The ferromagnetic exchange in $Ga_{1-x}Mn_xAs$ results from coupling between Mn ions at Ga sites ($Mn_{Ga}$) that is mediated by holes self-generated by $Mn_{Ga}$.[3] However, $Mn_{Ga}$ are partially compensated by other impurities, including Mn at interstitial sites ($Mn_I$).[4] $Mn_I$ are double donors and are thought to exhibit an antiferromagnetic exchange interaction with neighboring $Mn_{Ga}$[5] – making $Mn_I$ highly disruptive to ferromagnetism.

Annealing of $Ga_{1-x}Mn_xAs$ can greatly increase $T_C$[6] and the magnetization ($M$).[7,8] Understanding the mechanism of this annealing process is of utmost technological importance, in order to determine if $T_C$ can be pushed further towards room temperature. A recent study has shown that capping $Ga_{1-x}Mn_xAs$ thin films with GaAs suppresses any enhancement of $T_C$ or $M$ associated with annealing.[9] This corroborated other recent work suggesting that annealing causes $Mn_I$ to diffuse to the film surface, freeing additional $Mn_{Ga}$ to participate in the ferromagnetic exchange.[2,10,11]

In Ref. 11, we used polarized neutron reflectometry (PNR) to show that optimal annealing of an uncapped 100 nm $Ga_{1-x}Mn_xAs$ film ($x$ = 0.073) not only increased $T_C$ and $M$, but also changed the surface composition. Additionally, we found that the as-grown film had a pronounced gradient in $M$ that increased from the substrate to the surface – a feature that was significantly reduced after annealing. We have since seen these effects reproduced in thinner (but otherwise similar), uncapped $Ga_{1-x}Mn_xAs$ films.[12]

We have now expanded our study to probe the effects of annealing on the *depth-dependent* properties of $Ga_{1-x}Mn_xAs$ capped with GaAs. Using molecular-beam epitaxy, a $Ga_{1-x}Mn_xAs$ sample was prepared by first depositing a 160 nm GaAs buffer layer on a [001] GaAs substrate at a temperature of 580 °C, then cooling the substrate to 230 °C and adding another 2.7 nm GaAs buffer layer, before depositing a 100 nm film of $Ga_{1-x}Mn_xAs$, and then a 9 nm GaAs cap. Using x-ray diffraction, the $Mn_{Ga}$ concentration of the film was estimated to be $x = 0.076$.[13] This sample was cleaved, and one piece was annealed in $N_2$ for 1 h at 270 °C (nominally the same conditions as in Ref. 11), while another piece was left as-grown. These pieces were further cleaved, providing separate specimens for PNR and SQUID-based magnetometry.

The net $M$ of the samples, obtained using the magnetometer, is shown in Fig. 1. Fields were applied along a [110] direction. These measurements show that, in sharp contrast with uncapped samples, annealing does not improve the ferromagnetic properties (in agreement with Ref. 9). In fact, we observe that annealing is *detrimental* to the sample's ferromagnetic properties, as the low-field $T_C$ is reduced from 53 K to 40 K, and the high-field $M$ at $T = 13$ K drops from 23 emu·cm$^{-3}$ to 17 emu·cm$^{-3}$. Annealing of a similar uncapped $Ga_{1-x}Mn_xAs$ sample in the same oven at the same time as the capped sample resulted in a significant increase in $T_C$ (from 40 K to 90 K) - evidence that the GaAs cap is indeed responsible for ruining the beneficial effects of annealing.

PNR measurements were conducted using the NG-1 Reflectometer at the NIST Center for Neutron Research. A magnetic field of $H \approx 6.6$ kOe was applied in the plane of the sample along a [100] direction before cooling it to $T = 18$ K. Neutrons were spin-polarized either parallel or antiparallel to $H$, and were specularly reflected from the sample. The non spin-flip ($R_{++}$ and $R_{--}$) and spin-flip ($R_{+-}$ and $R_{-+}$) reflectivities were measured as functions of wavevector transfer $Q$. The data was corrected for $Q$-dependent sample illumination, and for instrumental background. The spin-flip scattering was minimal, and was used only to make polarization efficiency corrections to the data.

Figure 2 shows the corrected PNR data and fits in terms of spin asymmetry

$$SA = (R_{++} - R_{--}) / (R_{++} + R_{--}),  \quad (1)$$

which is a convenient quantity for gauging the sample magnetization parallel to $H$ at different length scales. The spin asymmetries for the as-grown and annealed samples are very similar, as the oscillations for both are "smeared" - indicative of magnetic roughness.[11] The amplitude of the lowest-$Q$ peak is larger for the as-grown sample, consistent with a slightly reduced net $M$ after annealing.

Depth-dependent magnetic and structural properties can be deduced by fitting PNR data with a scattering length density (SLD) model.[14,15] The SLD can be expressed as the sum of a chemical component[16] (dependent on the concentration of the constituent elements) and a magnetic component (proportional to $M$).[17] The fits to the $R_{++}$ and $R_{--}$ reflectivities are represented by the solid lines through the spin asymmetry data in Fig. 2, and were generated using *Reflpol* PNR fitting software.[18] The fits match the data well, and correspond to the SLD models shown in Fig. 3. Bracketed by GaAs on either side, the $Ga_{1-x}Mn_xAs$ film shows up clearly in each model, denoted by a region of decreased chemical SLD[19], and non-zero magnetic SLD. Uncertainty in the models' *net* magnetizations was reconciled by choosing models in which the integrated $M$ is consistent with that obtained from magnetometry measurements.

An interesting difference between the as-grown and annealed samples is a small increase in chemical roughness at the substrate/film interface for the annealed sample – a phenomenon we have also commonly observed for uncapped films. Otherwise, we see that annealing changes the depth profiles very little. Both $Ga_{1-x}Mn_xAs$ films feature a pronounced gradient in $M$ that extends over a thickness of approximately 500 Å ± 100 Å. (Note that --although the fit is not highly sensitive to the exact extent of this gradient -- the data unambiguously require that the model magnetization near the substrate be greatly depleted.) We therefore conclude that the reduction in net $M$ upon annealing occurs uniformly, and that annealing does not "smooth out" $M$ as we observe it to do for uncapped films. Since these $M$ gradients appear to be more prominent for films with lower $T_C$ and net $M$, it seems likely that they are correlated with increased $Mn_I$ concentration. However, for both films the chemical SLD changes little over the region of graded $M$ - implying that the *total* Mn concentration is relatively constant. Therefore, it is likely that the $M$ gradients are indicative of a non-uniform, depth-dependent ratio of $Mn_{Ga}$ to $Mn_I$, due to small, depth-dependent differences in growth temperature. If this is the case, a $M$ gradient is a "signature" of the $Mn_{Ga}/Mn_I$ ratio. The fact that this signature is unchanged upon annealing suggests that the cap prevents any large-scale redistribution of $Mn_I$. This result strongly supports an argument put forth in Ref. 2 and Ref. 9. These authors propose that as positively charged donor ions, $Mn_I$ are electrostatically prevented from diffusing into GaAs in large quantities due to formation of a p-n junction with negatively charged $Mn_{Ga}$ acceptor ions. Therefore, it seems that annealing of a capped sample provides enough energy to liberate $Mn_I$, but once junctions are formed, the $Mn_I$ has no preferential direction in which to travel! However, while it does not appear that large amounts of $Mn_I$ are vertically migrating during annealing, it is possible that a small number of $Mn_{Ga}$ do leave their lattice sites and form Mn clusters or MnAs inclusions – which could explain the observed drop in $T_C$ and net $M$.

In summary, we have observed that a GaAs capping layer not only eliminates the beneficial effects of annealing, but it also prevents annealing from altering the depth-dependence of the magnetization of $Ga_{1-x}Mn_xAs$. These results lend further support to a model of annealing of uncapped $Ga_{1-x}Mn_xAs$ in which the added energy pries $Mn_I$ from the lattice, where it then falls into an electrostatic potential that draws it towards the free surface. Additionally, we see evidence that a non-uniform magnetization is a common feature of $Ga_{1-x}Mn_xAs$ growth – a factor that may warrant consideration for potential device applications.

This work is supported by NSF Grant No. DMR-013819. Special thanks go to Suzanne te Velthuis of Argonne National Laboratory for valuable discussions, and to Paul Kienzle of NIST for development of and assistance with *Reflpol* PNR software.

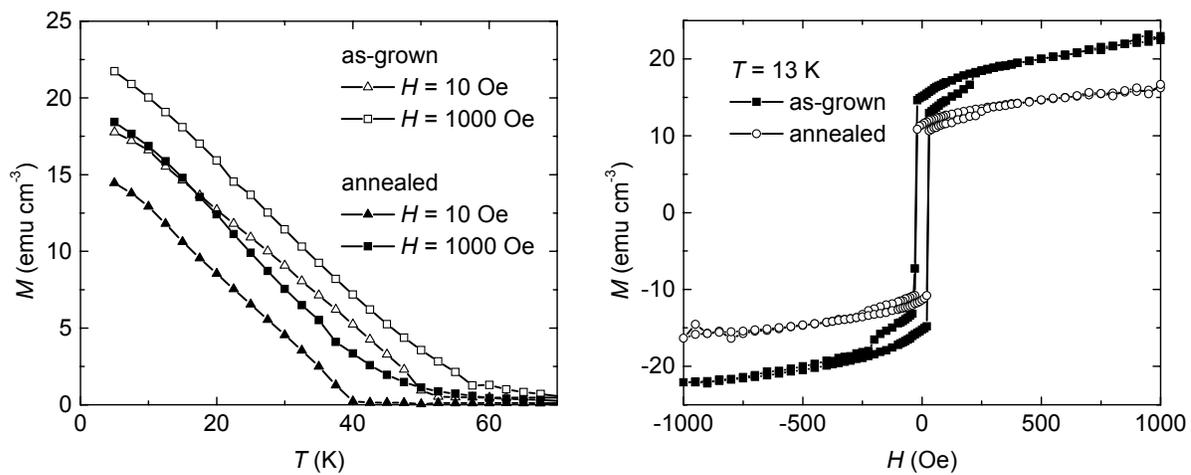

**B. J. Kirby, Figure 1.** SQUID-based magnetometry results showing the net magnetizations of the as-grown and annealed films.

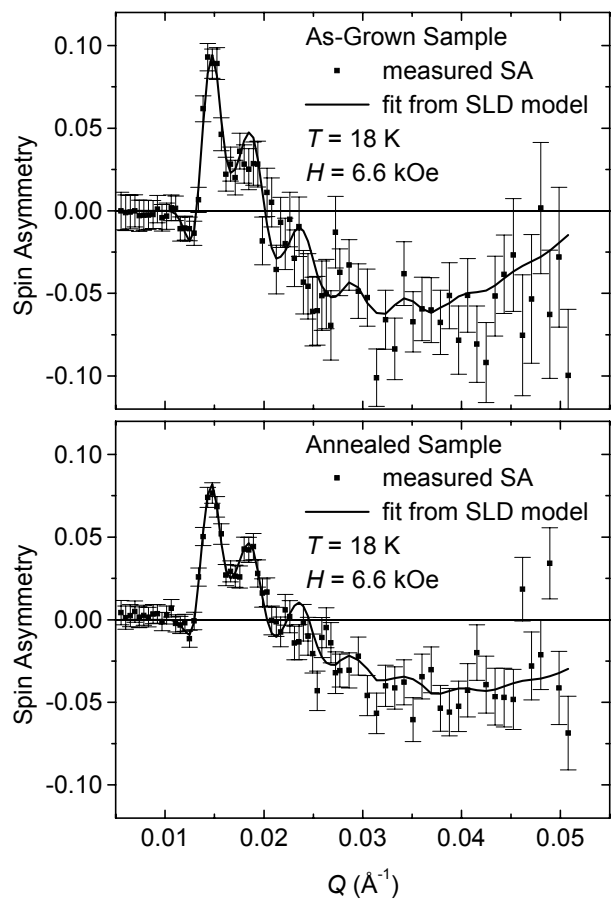

**B. J. Kirby, Figure 2. PNR data and fits displayed as spin asymmetry (defined in the text).**

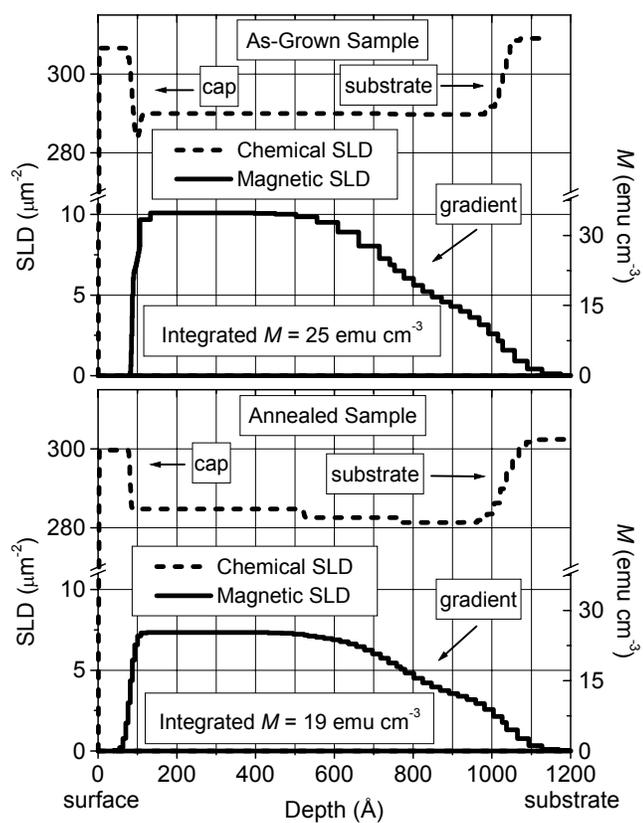

**B. J. Kirby, Figure 3.** Scattering length density models used to fit the data in Fig. 2.

---

[13] G. M. Schott, W. Fashinger, and L. W. Molenkamp, *Appl. Phys. Lett.* **79**, 1807 (2001).

[14] C. F. Majkrzak, *Physica B* **173**, 75 (1991).

[15] J. F. Ankner, C. F. Majkrzak, and H. Homma, *J. Appl. Phys.* **73**, 6436 (1993).

[16] Also called "nuclear" SLD.

[17] G. P. Felcher, *Phys. Rev. B* **24**, 1595 (1981).

[18] Information can be found at http://www.ncnr.nist.gov/programs/reflect/.

[19] Difference in substrate chemical SLD between the two films is due to differences in reflectometer alignment. These alignment differences do not significantly affect the magnetic SLD, or the distribution of chemical SLD.